# 5.0 µm Emitting Interband Cascade Lasers with Superlattice and Bulk AlGaAsSb Claddings

B. Petrović,[1] A. Bader,[1] J. Nauschütz,[2] T. Sato,[3] S. Birner,[3] R. Weih,[2] F. Hartmann,[1] and S. Höfling[1]

Electronic mail: borislav.petrovic@physik.uni-wuerzburg.de
fabian.hartmann@physik.uni-wuerzburg.de

[1]*Julius-Maximilians-Universität Würzburg, Physikalisches Institut, Lehrstuhl für Technische Physik, Am Hubland, 97074 Würzburg, Germany*

[2]*nanoplus Advanced Photonics Gerbrunn GmbH, Oberer Kirschberg 4, 97218 Gerbrunn, Germany*

[3]*nextnano GmbH, Konrad-Zuse-Platz 8, 81829 München, Germany*

We present a comparison between interband cascade lasers (ICLs) with a 6-stage active region emitting at 5 µm with AlSb/InAs superlattice claddings and with bulk $Al_{0.85}Ga_{0.15}As_{0.07}Sb_{0.93}$ claddings. Utilizing bulk AlGaAsSb claddings with their lower refractive index compared to the more commonly used AlSb/InAs superlattice claddings, the mode-confinement in the active region increases by 14.4 % resulting in an improvement of the lasing threshold current density. For broad area laser and under pulsed excitation, the ICL with AlGaAsSb claddings shows a lower threshold current density of $J_{th} = 396\ A/cm^2$ compared to $J_{th} = 521\ A/cm^2$ of the ICL with superlattice claddings. Additionally, a higher characteristic temperature was obtained for the ICL with bulk claddings. Emission in pulsed operation is observed up to 65 ºC.

## I. INTRODUCTION

Over the years, interband cascade lasers (ICLs[1]) have been established as promising mid-infrared light sources for gas tracing applications finding its use in clinical breath analysis[2], atmosphere exploration[3], industrial process control[4] and missile tracking[5]. Similarly to quantum cascade lasers (QCLs), the cascading architecture of ICLs has lower parasitic voltage drops compared to parallel configuration of multiple quantum wells in laser diodes[6]. Due to interband transitions and long upper state lifetimes, ICLs operate at lower threshold current densities and lower threshold power densities, especially in the 3-6 µm wavelength range. They are energy efficient light sources ideally suited for portable applications[7]. When the first continuous wave (cw) operation at room temperature was demonstrated in 2008[8], use of lightly doped separate confinement layers along with superlattice claddings were introduced and set as state of the art ever since[6-7].

AlSb/InAs superlattices are commonly employed in claddings of ICLs due to the possibility of strain-compensated growth on a GaSb substrate[6-7]. A significant work has been done recently to investigate alternative cladding designs with the aim to reduce the refractive index while maintaining and enhancing the thermal conductivity[9-13]. Although AlAs and AlSb have the lowest refractive indices out of III-V binaries in the mid-infrared wavelength region, a lattice matched molecular beam epitaxy (MBE) growth of the $AlAs_{0.08}Sb_{0.92}$ ternary alloy as the cladding material is not likely to be sustainable due to excessive oxidation for materials with high Al content[9]. A higher oxidation stability is obtained by adding a low amount of Ga to the alloy, although it will increase the refractive index. As reported in Ref. 9 for ICLs emitting at 3.5 μm, ICL employing bulk AlGaAsSb claddings are promising and can potentially outperform ICLs with AlSb/InAs superlattice claddings. The reported substrate temperature during the growth of the AlGaAsSb layers was 500 ºC. A similar threshold current density at approximately the same emission wavelength was obtained in Ref. 10 where the growth temperature optimization of the AlGaAsSb layers was reported. In this letter, we realize and compare ICLs emitting at 5 μm and contrast the operation of ICLs with InAs/AlSb superlattice and AlGaAsSb bulk claddings. Obtaining low threshold current densities at longer wavelengths, i.e. above the sweet spot operation of ICLs (~3.5 μm), becomes technologically challenging since the wavefunction overlap in the W-type quantum well lasing transition decreases and free carrier absorption losses increase. In this paper we present ICLs with active region designed to emit at 5 μm and evaluate the advantages and disadvantages of quaternary bulk AlGaAsSb claddings for ICLs.

## II. MODELING

Table I shows an overview of the refractive indices $n_r$ and thermal conductivities $\kappa$ of different cladding designs for materials lattice matched to GaSb.

|  | AlSb/InAs | AlGaAlSb | n$^+$-InAsSb |
|---|---|---|---|
| $n_r$ | 3.39 | 3.30 | ≈2.70 |
| $\kappa$ (W/m·K) | ≈3 | ≈7 | ≈15 |

Table I: Refractive indices and thermal conductivities of AlSb/InAs superlattice claddings, bulk $Al_{0.85}Ga_{0.15}As_{0.07}Sb_{0.93}$ claddings, and plasmon-enhanced highly doped $InAs_{0.915}Sb_{0.085}$ claddings.

Two ICL structures with the same 6-stage active region and geometry but different cladding designs are depicted in Figure 1. The ICL with superlattice (SL) AlSb/InAs claddings is depicted in Figure 1 (a) and ICL with bulk $Al_{0.85}Ga_{0.15}As_{0.07}Sb_{0.93}$ claddings in Figure 1 (b).

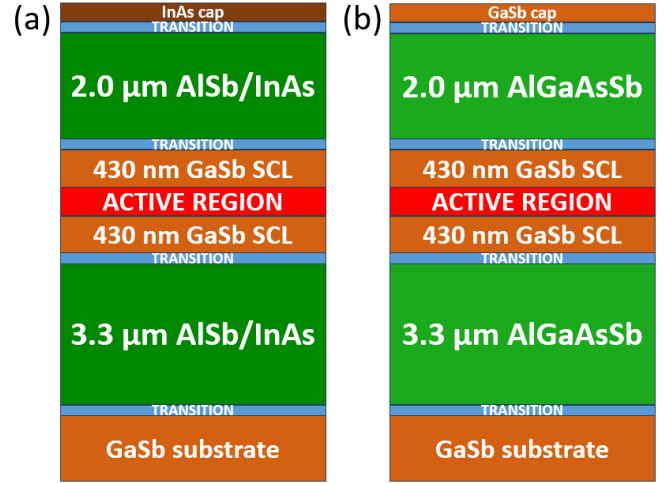

FIG. 1. (a) Schematic layout of the ICLs with SL cladding in (a) and with AlGaAsSb bulk cladding in (b).

$Al_{0.85}Ga_{0.15}As_{0.07}Sb_{0.93}$ has a lower refractive index $n_{r\,AlGaAsSb} = 3.30$[10] compared to $n_{r\,AlSb/InAs} = 3.39$ of InAs/AlSb SLs[9,14-15]. In addition, a higher thermal conductivity $\kappa$ of the bulk AlGaAsSb[10,12,13] was reported. While the lower refractive index provides a better mode-confinement of the active region, the higher thermal conductivity could be beneficial for continuous wave operation. In the third column, refractive index[16] and thermal conductivity[13] of highly doped ($1 \cdot 10^{19} cm^{-3}$) plasmon-enhanced $InAs_{0.915}Sb_{0.085}$ is presented as another cladding alternative for ICLs grown on GaSb substrates. Such cladding design has been successfully implemented in ICLs grown on InAs substrates[17-19] and readily transferred to ICLs grown on GaSb substrates[20-22].

The confinement factors of the two ICL structures were calculated by a Helmholtz wave equation solver. Figure 2 (a) shows the refractive index profiles $(n_r)$[14-15] and relative mode intensities (R.I.) of the two ICLs. Figure 2 (b) and (c) show the optical mode profiles with distributions of confinement factors Γ for the ICLs with SL and bulk AlGaAsSb claddings. Optical confinement in the active region of the ICL with bulk claddings is 14.4 % higher in comparison to ICL with SL claddings. In addition, confinement in the claddings is reduced by 17.8 %. Considering doping levels ($N$), effective masses[23] ($m_e^*$), mobilities[16,24-27] ($\mu_e$) listed in Table II, one can calculate free carrier absorption (FCA) losses in active region (AR) SCLs and both cladding materials, Eq. (1)

(1) $\alpha = \frac{e^3}{4\pi^2 c^3 \varepsilon_0} \cdot \frac{N\lambda^2}{\mu_e n_r m_e^{*2}}$

Considering the doping of the cladding layers linearly increases from $1 \cdot 10^{17} cm^{-3}$ to $1 \cdot 10^{18} cm^{-3}$ with distance from the active region, the FCA is also linearly dependent on the coordinate along the direction of the growth (x). From distribution of confinement factors, overall FCA losses can be evaluated in both SL and bulk cladding-based ICLs, with Eq. (2):

(2) $FCA = FCA(cladd, SCL, AR) =$

$\int \alpha_{cladd.}(x) d\Gamma_{cladd.}(x) + \alpha_{SCL}\Gamma_{SCL} + \alpha_{AR}\Gamma_{AR}$.

|   | SCL | SL cladd | Bulk cladd. | AR |
|---|---|---|---|---|
| $N$ (cm$^{-3}$) | 6E16 | 1E17-1E18 | 1E17-1E18 | 4E18 |
| $m_e^*$ (m$_0$) | 0.039 | 0.081 | 0.126 | 0.067 |
| $\mu_e$ (cm$^2$/Vs) | 2800 | 10.000 | 640 | 4500 |
| $FCA_{SL}$ (cm$^{-1}$) | 2.4 | 0.5 | / | 2.7 |
| $FCA_{Bulk}$ (cm$^{-1}$) | 2.6 | / | 2.3 | 3.1 |

Table II: Doping levels, effective masses, mobilities and free career absorption (FCA) losses in GaSb SCL, AlSb/InAs SL, bulk $Al_{0.85}Ga_{0.15}As_{0.07}Sb_{0.93}$ claddings and the active region.

The calculated overall FCA losses in the ICL with AlGaAsSb claddings was $\alpha = 8.0$ cm$^{-1}$ and is significantly higher than in the ICL with the standard design based on SL claddings with FCA losses of $\alpha = 5.6$ cm$^{-1}$ This suggests a tradeoff between the optical confinement improvement and enhanced free carrier absorption loss for the bulk quaternary cladding. A technological advantage of bulk quaternary cladding is the reduced shutter cycle operation in contrast to the over 1000 shutter cycles needed for the growth of SL claddings. However, growth of AlGaAsSb quaternary is also demanding due to appearance of amorphous surface defects caused by heavily n-type Te doping of AlGaAsSb layers[28]. The active region (AR) is designed for a laser emission wavelength of 5 μm and was performed using the commercial software nextnano[29]. The AR consists of 6 stages and employs InAs/Ga$_{0.6}$In$_{0.4}$Sb/InAs W-shaped quantum wells.

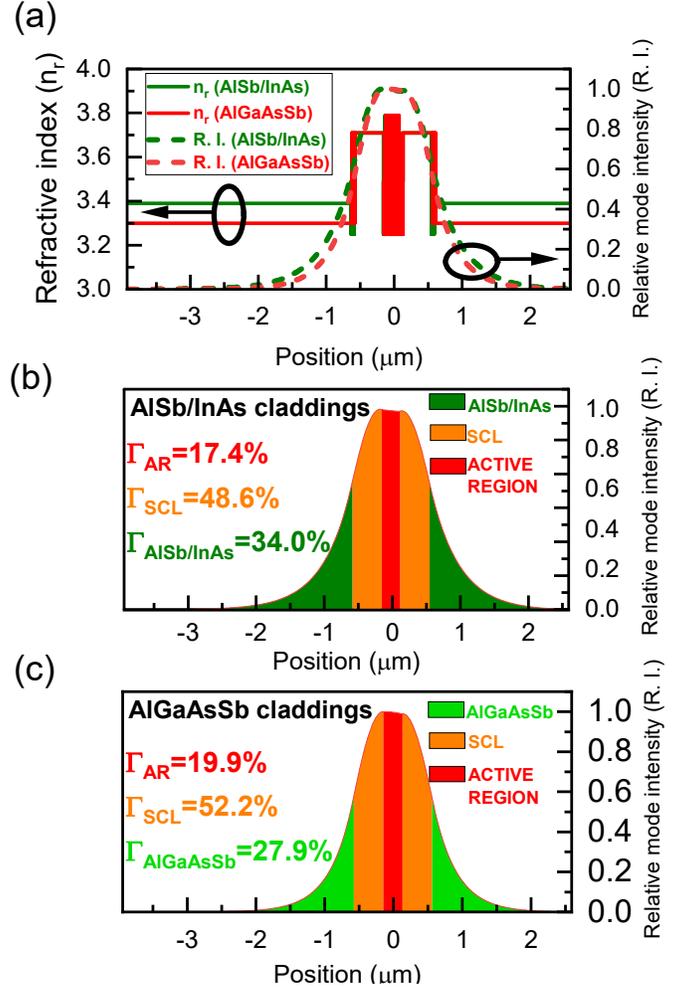

FIG. 2. (a) Refractive index profiles and relative mode intensities along the growth direction of two ICLs with the same active region design but different cladding designs: Superlattice (SL) InAs/AlSb claddings (green) and bulk AlGaAsSb claddings (red) (b) Optical mode confinement distribution along the growth direction of the ICL with SL claddings (c) Optical mode confinement distribution along the growth direction of the ICL with bulk AlGaAsSb claddings.

## III. EXPERIMENT

The ICLs were grown by MBE equipped with effusion cells for the group III elements, silicon, and tellurium as n-type dopants and two valved cracker cells for arsenic and antimony. The substrates used for both ICLs were 2'' GaSb, n-type doped with Te ($1 \cdot 10^{18} cm^{-3}$). A pyrometer was used for monitoring the substrate temperature during the growth. Prior to the growth, an oxide desorption step was performed under Sb flux at 560 ºC for 3 minutes. The growth starts with a 200 nm thick n-type doped GaSb:Te buffer layer ($1 \cdot 10^{18}$ cm$^{-3}$) grown at a substrate temperature of 500 ºC. Subsequently, for the ICL with SL claddings, the substrate temperature was ramped down to 450 ºC during the growth of the bottom claddings. 3.3 and 2 μm long claddings were grown as a strain compensated superlattice of 1120 periods of 2.43 nm thick InAs and 2.30 nm thick AlSb layers. InAs layers of the SL claddings were n-type doped with Si with a linearly decreasing doping concentration towards the active region ($1 \cdot 10^{18}$ cm$^{-3}$ to $1 \cdot 10^{17}$ cm$^{-3}$) to minimize optical losses. The substrate temperature was increased back to 500 ºC for the growth of the bottom separate confinement layer (SCL). During the growth of the ICL with bulk claddings, after the buffer, the temperature was kept at 500 ºC throughout the growth of $Al_{0.85}Ga_{0.15}As_{0.07}Sb_{0.93}$ cladding layer and the bottom SCL. The AlGaAsSb claddings were n-doped by Te with the same dopant ramp as for the ICL with SL claddings. Following the growth of the bottom SCL for both ICLs the substrate temperature was decreased to 450 ºC for the growth of the 6-stage active region. Once the active region was grown, the substrate temperature remained unchanged at 450 ºC. Separate confinement layers in both ICLs were 430 nm thick to increase the confinement in the active region and lightly n-type doped with Te ($6 \cdot 10^{16}$ cm$^{-3}$) to reduce the optical losses. The active W-QW is composed of **AlSb**/*InAs*/$Ga_{0.6}In_{0.4}Sb$/*InAs*/**AlSb** with thicknesses of (**2.5**/*2.23*/2.55/*1.75*/**1.0**) nm. The hole and electron injector are composed of 2.8 nm GaSb/1.0 nm AlSb/4.8 nm GaSb and 2.5 nm AlSb/4.6 nm InAs/1.2 nm AlSb/3.4 nm InAs/1.2 nm AlSb/2.6 nm InAs/1.2 nm AlSb/2.15 nm InAs respectively. Four InAs wells of the electron injectors were n-type doped to $4 \cdot 10^{18}$ cm$^{-3}$ for carrier rebalancing[30]. Due to the band offsets between the cladding layers and SCLs, transition layers composed of $Al_{0.85}Ga_{0.15}As_{0.07}Sb_{0.93}$/GaSb were implemented. Similarly, AlSb/InAs transition layers were incorporated between the SCLs and the active region. The ICL with SL architecture was capped with 25 nm of InAs doped to $2 \cdot 10^{19}$ cm$^{-3}$. The ICL with AlGaAsSb claddings was capped with GaSb with $3 \cdot 10^{18}$ cm$^{-3}$. The GaSb cap was used instead of InAs to reduce the conduction band offset between the bulk claddings and cap material to minimize their thickness.

Figure 3 (a) displays a high-resolution X-ray diffraction (HR-XRD) scan of the ICL with SL claddings. The GaSb substrate and $0^{th}$ order SL reflex labeled in the figure show a good strain compensation with a mismatch of only 560 ppm. Higher order ($1^{st} - 3^{rd}$) satellite peaks of the SL claddings, active region peaks (AR) and InAs peak of the cap layer are also shown. Figure 3 (b) presents the HR-XRD scan of the ICL with AlGaAsSb claddings, highlighting almost perfect lattice matching of the claddings to the GaSb substrate with a mismatch as small as 340 ppm. Active region peaks show a good quality of the grown ICLs of both SL and AlGaAsSb cladding architectures indicated by visible peaks up to $17^{th}$ and $19^{th}$ order, respectively.

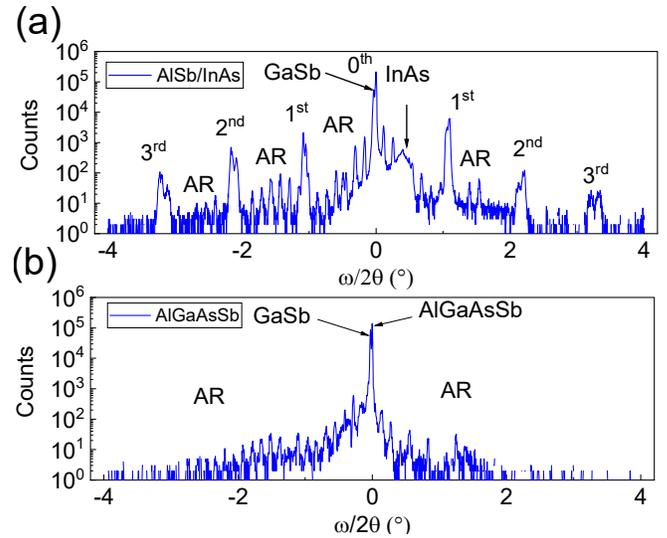

FIG. 3. (a) X-ray diffractogram of the ICL with SL claddings indicating strain compensated growth (b) X-ray diffractogram of the ICL with AlGaAsSb bulk claddings indicating lattice matched growth

Deep etched broad area (BA) devices were fabricated for characterization in pulsed mode. For processing of BA devices, standard photolithography was used to define 100 μm wide ridges. In the wet etching process a mixture of $H_2O/H_3PO_4/H_2O_2/$ $HOC(CO_2H)(CH_2CO_2H)_2$ was used to etch through the active region and the bottom SCL. Afterwards, a Ti/Pt/Au top contact was deposited, and AuGe/Ni/Au layers were evaporated at the substrate side as a back contact. The fabricated sample was cleaved into 2 mm long laser bars and measured epilayer side up on a copper heat sink.

## IV. RESULTS AND DISCUSSION

Pulsed measurements were performed under a repetition rate of 1 kHz and the applied current pulse width was 500 ns to minimize Joule heating. Figure 4 (a) displays electro-optical characteristics of the BA ICLs operated in pulsed mode at room temperature (25 ºC). At room temperature the ICL with SL claddings shows a threshold current density of $J_{th} = 521$ A/cm$^2$ threshold power density of $P_{th} = 2.03$ kW/cm$^2$, threshold voltage of $V_{th} = 3.9$ V and a voltage efficiency of $\eta_V = 38.1$ %, whereas for the ICL with bulk AlGaAsSb claddings the corresponding figures of merit are $J_{th} = 396$ A/cm$^2$, $P_{th} = 2.65$ kW/cm$^2$, $V_{th} = 6.7$ V and $\eta_V = 22.2$ %. In the range 0-3.6 V, a presence of high leakage currents in both ICLs with SL and AlGaAsSb claddings is evident. Assuming ohmic behavior of sidewalls responsible for the leakage, electrical characteristics of the leakage components of the currents are extrapolated to higher voltages. We would like to note that the leakage current contribution is an estimate based on the BA process data. A more reliable estimation though, would involve analysing ICLs of different perimeter to area ratios (P/A-ratio) and to extract the leakage current component via extrapolation of the P/A-ratio[31]. In Figure 4 (b) the linear fits of the leakage currents are shown in respect to measured electrical characteristics of the ICLs. Leakage currents at threshold voltages (labeled in blue) correspond to 154 A/cm$^2$ and 168 A/cm$^2$ for ICLs with SL and AlGaAsSb claddings respectively. Considering that the leakage currents at the respective thresholds are approximately the same, they are unlikely to contribute to the difference in threshold current densities of the two ICLs. Nevertheless, the high leakage currents contribute to above average threshold current densities[21,32-33] (≈300 A/cm$^2$) and consequently above state-of-the-art threshold voltages (≈ 2 V)[34] and threshold power densities (≈0.8-1.5 kW/cm$^2$)[6]. The threshold current densities are significantly higher than values reported for ICL with plasmon-enhanced claddings lasing at $\lambda = 5.17$ μm[35] ($J_{th} = 306$ A/cm$^2$) and $\lambda = 4.3 - 4.8$ μm[36] ($J_{th} = 252$ A/cm$^2$).

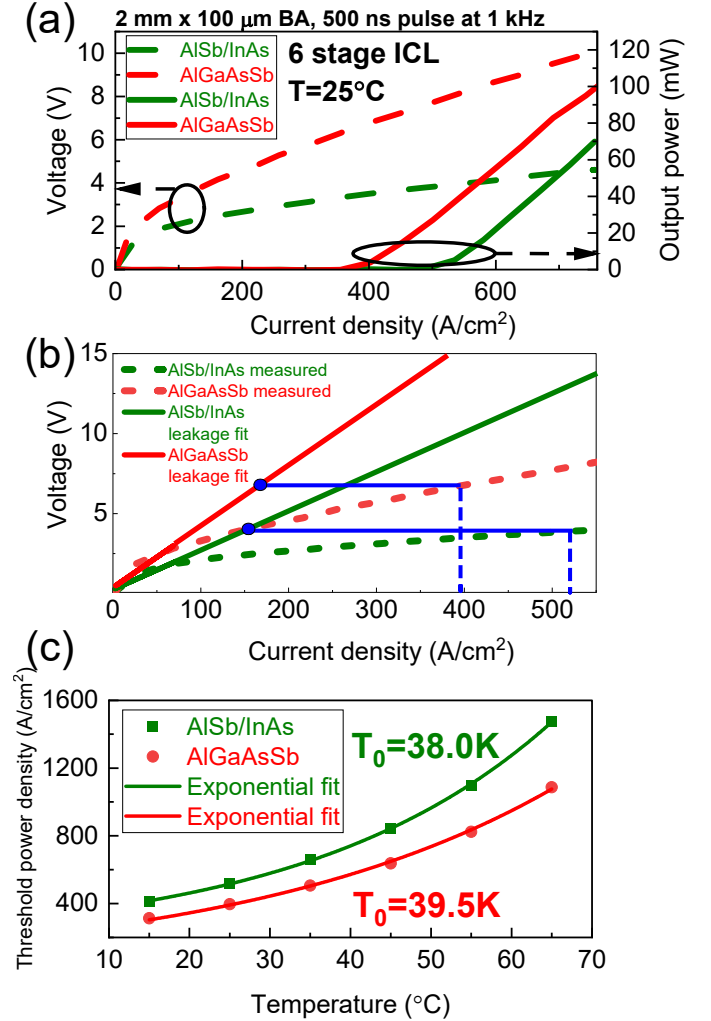

FIG. 4. (a) Pulsed light-current and light-voltage curves at room temperature for a 2 mm long and 100 μm wide BA ICLs with SL and lGaAsSb bulk claddings. (b) Voltage characteristics of the BA ICLs with SL and AlGaAsSb claddings with linear fits of their leakage components and labeled leakage currents at respective threshold voltages. (c) Temperature dependencies of the threshold current densities and characteristic temperatures of ICLs with SL claddings $T_0$ =38.0 K, and AlGaAsSb bulk claddings $T_0$ =39.5 K.

The papers highlight use of such cladding designs as their performance emphasizes their remarkable benefits. The voltage efficiencies are significantly lower than state-of-the-art value of $\eta_V \approx 60\%$[35] and the reported values of $\eta_V = 63\%$[35] and $\eta_V = 80\%$[36] which is mainly a consequence of high threshold current densities. The smaller threshold current density of the ICL with the bulk claddings suggests that the higher mode-confinement in the active region has a higher influence than free carrier absorption losses. In addition, the calculated free carrier absorption losses might be overestimated. The smaller voltage efficiency of the ICL with bulk cladding is due to parasitic voltage drops across the transition layers connecting AlGaAsSb claddings with the rest of the heterostructure. The differential resistance of the ICL with SL claddings of 2 Ω is in agreement with typical 1-2 Ω[9]. However, the high value of 5.4 Ω measured for the ICL with AlGaAsSb claddings is caused by low doping of the cap and the transition layers between GaSb and AlGaAsSb layers. Due to only 15% of Ga content in the quaternary, the contrast between the growth rates of the alternating $Al_{0.85}Ga_{0.15}As_{0.07}Sb_{0.93}$ and GaSb layers is high. As the doping cell temperature cannot be changed rapidly, in order to avoid overdoping GaSb layers, doping levels of AlGaAsSb layers in the transition regions are relatively low. Remarkably higher series resistance in ICL with bulk claddings contributes to its higher threshold voltage. Figure 4 (c) depicts the temperature dependencies of threshold current densities for both ICLs and shows that the characteristic temperatures are comparable. The characteristic temperatures compare well with previously reported values at similar wavelengths[21-22,35,37]. The measured figures of merit and number of cascade stages (N) with respective previously reported values[6,7,17,34,36,37] at room temperature and at similar wavelengths are listed in the Table III.

|  | SL | AlGaAsSb | Literature |
| --- | --- | --- | --- |
| $J_{th}$ (A/cm²) | 521 | 396 | 247-500 |
| $P_{th}$ (kW/cm²) | 2.03 | 2.65 | 0.8-1.5 |
| $V_{th}$ (V) | 3.9 | 6.7 | ≈2 |
| N | 6 | 6 | 5-12 |
| Substrate type | n-GaSb | n-GaSb | n-GaSb/InAs |
| $T_{max}$ (°C) | 65 | 65 | ≤106 |
| $T_0$ (K) | 38.0 | 39.5 | 35-57 |

Table III: Figures of merit of the ICLs with SL and AlGaAsSb claddings with values reported in literature for ICLs at room temperature and wavelengths of approximately 5 μm.

In addition to 2 mm long laser bars, two series of BA devices of lengths 1.4 mm, 1.6 mm and 1.8 mm have been fabricated and measured in pulsed operation to extract the absorption losses. Figures 5 (a) and 5 (b) depict current density-optical power characteristics of BA ICLs with SL and bulk AlGaAsSb claddings, respectively. At fixed current density, the output power is increasing with length of the cavity. In the subplots, the corresponding external differential quantum efficiencies (EDQEs) are presented. Slopes of the optical power-current characteristics are aproximated to be linear in the range 0-1200 A/cm² hence the EDQEs remain constant after reaching threshold. In the Figure 5 (c) reciprocal values of EDQEs at current density of 1000 A/cm² are shown for different cavity lengths. From this dependence, assuming facet reflectivity of R=0.31 we can extract the internal efficiency $\eta_i$ and absorption losses $\alpha_w$ of the waveguide (active region, claddings and SCLs), according to Eq. (3)[38]

(3) $\quad \frac{1}{EDQE} = \frac{1}{\eta_i}(1 + \frac{\alpha_w L}{\ln\frac{1}{R}})$

The obtained internal efficiency and waveguide absorption loss for the ICL with SL claddings are $\eta_i = 0.177$ and $\alpha_w = 4.8$ cm$^{-1}$, whereas for the ICL with AlGaAsSb claddings the values are $\eta_i = 0.203$ and $\alpha_w = 7.3$ cm$^{-1}$ respectively. The ICLs have shown similar internal efficiencies but significantly below values of state-of-the-art[6,38] due to high threshold current densities. Conversely, there is a significant difference in the absorption losses. The absorption losses in the ICLs with SL and bulk claddings are slightly lower than calculated (5.6 cm$^{-1}$ and 8.0 cm$^{-1}$). The values are reasonably close, however small deviations could be explained by fiting approximations of the electro-optical characteristics and accuracy of theoretical values of electron mobilities.

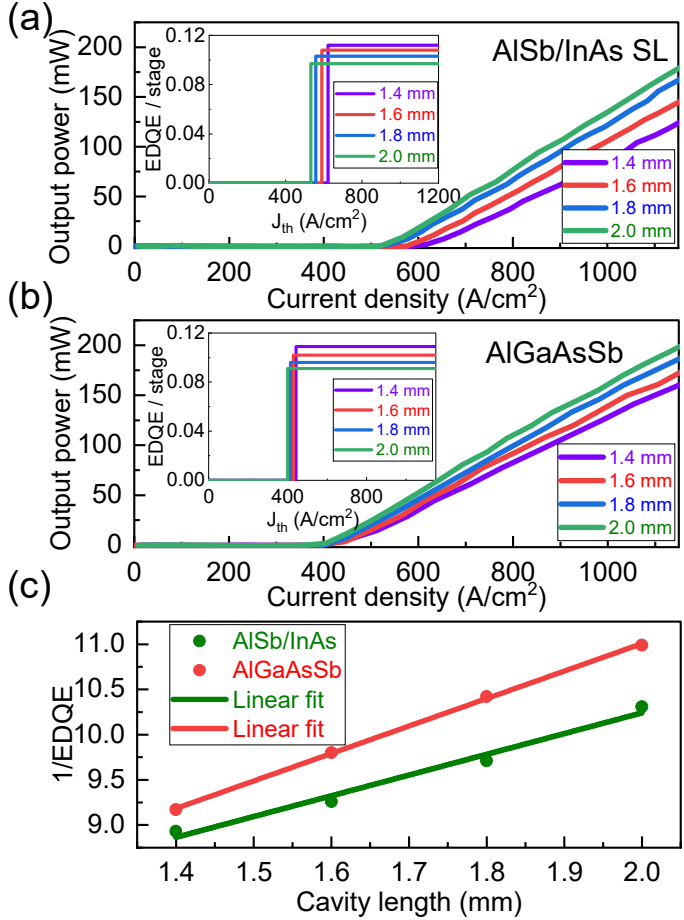

FIG. 5. (a) Pulsed light-current curves at room temperature for 1.4 mm, 1.6 mm, 1.8 mm and 2.0 mm long and 100 µm wide BA ICLs with SL claddings. In the subplot the corresponding external differential quantum efficiencies (EDQEs) are shown. (b) Pulsed light-current curves at room temperature for 1.4 mm, 1.6 mm, 1.8 mm and 2.0 mm long and 100 µm wide BA ICLs with AlGaAsSb claddings. In the subplot the corresponding EDQEs are shown. (c) Dependences of inverse EDQEs of the ICLs with SL and AlGaAsSb claddings on the cavity lengths.

Figures 6 (a) and (b) show the lasing spectra at three different heat sink temperatures of the ICL with SL and AlGaAsSb bulk claddings, respectively. Central wavelengths of the two ICLs are $\lambda_{SL} = 4.93$ µm and $\lambda_{AlGaAsSb} = 5.04$ µm. Longer wavelengths contribute to higher intervalence band absorption for suboptimal 2.50 nm thick GaInSb hole quantum well according to Ref 37, which might also contribute to higher threshold current densities. However, the ICL with bulk claddings has shown a lower threshold current density than the ICL with SL claddings suggesting the benefits coming from the use of AlGaAsSb. The temperature-induced wavelength shift is 2.25 nm/°C for both ICLs.

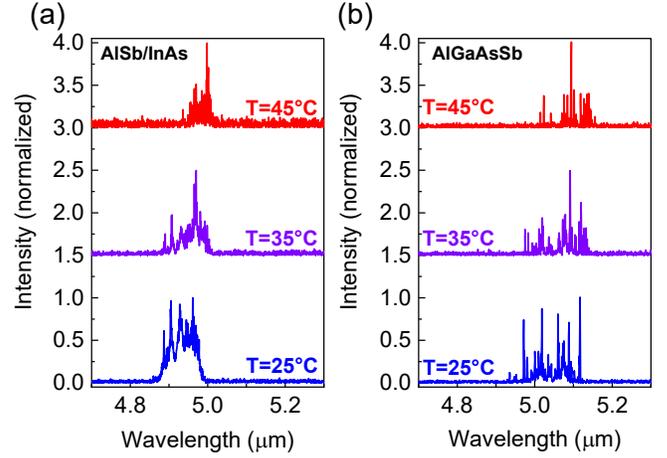

FIG. 6. Spectra at 25 ºC, 35 ºC and 45 ºC of BA (2 mm x 100 µm) devices for the ICL with SL claddings (in (a)) and the ICL with bulk AlGaAsSb claddings (in (b)). The central wavelengths at room temperature are 4.93 µm and 5.04 µm respectively.

## V. SUMMARY AND CONCLUSIONS

In summary, we have compared two ICLs with the same active region design but different cladding design. A reduced threshold current density and higher characteristic temperature was obtained for the sample based on the bulk quaternary cladding design. On the contrary, the bulk design has shown higher threshold power density and voltage efficiency. This is likely a consequence of high series resistance coming from suboptimal design and doping of the transition layers and the cap layer. A possible improvement of their figures of merit could still be made by further optimizing growth conditions and transition layers. For instance, a highly doped InAs cap can be used instead of GaSb and InAs/AlSb instead of AlGaAsSb/GaSb transition layers could be a good alternative to avoid the low doping influenced by different rates of growth of the same material in adjacent layers. Another possible solution would involve use of MBE systems with two Ga cells. Different Ga-cell temperatures would enable growth of the adjacent $Al_{0.85}Ga_{0.15}As_{0.07}Sb_{0.93}$ and GaSb layers with the same rates of growth. This would avoid disparity in the doping levels.

Taking the previous into account, we conclude advantages of use bulk claddings in terms of threshold current density and the optical confinement, however at this stage the ultimate superiority of such cladding design cannot be resolved.

## ACKNOWLEDGEMENTS

We are grateful to European Union's Horizon 2020 research and innovation programme under the Marie Skłodowska-Curie grant agreement no 956548 (QUANTIMONY) for financial support of this work. We also thank S. Estevam for sample processing and preparation.

## AUTHOR DECLARATIONS

**Conflict of interest**

The authors declare no conflicts to disclose.

## DATA AVAILABILITY

The data that support the findings of the study are available from the corresponding author upon reasonable request.